\documentclass[twocolumn,english,showpacs,aip,cha,reprint]{revtex4-1}
\usepackage[T1]{fontenc}
\usepackage[latin9]{inputenc}
\usepackage{babel}
\usepackage{amsmath}
\usepackage{amssymb}
\usepackage{graphicx}
\usepackage{esint}
\usepackage[unicode=true,pdfusetitle,
 bookmarks=true,bookmarksnumbered=false,bookmarksopen=false,
 breaklinks=false,pdfborder={0 0 1},backref=false,colorlinks=false]
 {hyperref}
\usepackage{breakurl}

\makeatletter
 
 \@ifundefined{textcolor}{}
 {%
   \definecolor{BLACK}{gray}{0}
   \definecolor{WHITE}{gray}{1}
   \definecolor{RED}{rgb}{1,0,0}
   \definecolor{GREEN}{rgb}{0,1,0}
   \definecolor{BLUE}{rgb}{0,0,1}
   \definecolor{CYAN}{cmyk}{1,0,0,0}
   \definecolor{MAGENTA}{cmyk}{0,1,0,0}
   \definecolor{YELLOW}{cmyk}{0,0,1,0}
 }

\usepackage{babel}

\makeatother

\begin{document}

\title{Analytical approximations for spiral waves}

\author{Jakob Löber}

\email{jakob@physik.tu-berlin.de}

\selectlanguage{english}%

\author{Harald Engel}

\affiliation{Institut für Theoretische Physik, Technische Universität Berlin,
Hardenbergstrasse 36, 10623 Berlin, Germany}


\begin{abstract}
We propose a non-perturbative attempt to solve the kinematic equations
for spiral waves in excitable media. From the eikonal equation for
the wave front we derive an implicit analytical relation between rotation
frequency $\Omega$ and core radius $R_{0}$. For free, rigidly rotating
spiral waves our analytical prediction is in good agreement with numerical
solutions of the linear eikonal equation not only for very large but
also for intermediate and small values of the core radius. An equivalent
$\Omega\left(R_{+}\right)$ dependence improves the result by Keener
and Tyson for spiral waves pinned to a circular defect with radius
$R_{+}$ with Neumann boundaries at the periphery. Simultaneously,
analytical approximations for the shape of free and pinned spirals
are given. We discuss the reasons why the ansatz fails to correctly
describe the result for the dependence of the rotation frequency on
the excitability of the medium.
\end{abstract}
\maketitle

\section{Introduction}

Spiral waves are a common occurrence in excitable media. They have
been observed in systems as diverse as catalytic surface reactions
\cite{jakubith1990spatiotemporal}, the Belousov-Zhabotinsky chemical
reactions \cite{winfree1972spiral,zhabotinsky1973autowave} and social
amoeba colonies \cite{gerisch1971periodische}. They play an important
role as pathological states of action potential propagation in cardiac
tissue and are thought to be the precursor of ventricular fibrillation
\cite{fenton2002multiple}.

\indent In the simplest case, a free spiral rotates rigidly with
a frequency $\omega$ while its tip describes a circular trajectory
with radius $r_{0},$ called the spiral core radius. From experiments
and numerical simulations it is well-known that spiral waves select
their own, unique asymptotic wave shape and rotation frequency. Thus,
independently on the method of initiation, coexisting (and non-interacting)
spiral waves in a spatially uniform excitable or oscillatory medium
exhibit the same wave length, core radius and rotation frequency after
all transients have died out. Exceptions to this rule are known only
for media with anomalous dispersion of periodic wave trains \cite{Winfree1991Alternative}.

\indent A theoretical description aims to understand the underlying
selection principle determining shape, rotation frequency and core
radius of spiral waves. One successful approach is the free boundary
or kinematic approach \cite{zykov1987simulation,tyson1988singular}
which reduces the nonlinear reaction-diffusion equations to simpler
equations describing the motion of interfaces separating excited from
resting states. In contrast to earlier works, which mapped wave front
and wave back onto each other \cite{zykov1987simulation,davydov1991kinematics,mikhailov1994complex}\emph{,
}it became clear that one has to solve\emph{ }equations for both the
front and back interface of a spiral to yield a self-consistent solution
for $\omega$ and $r_{0}$.

\indent Within the free-boundary approach, the pattern selection
problem for wave segments \cite{zykov2005wave}, which are unstable
solutions of the reaction-diffusion system, has been solved. These
patterns undergo translational motion in an unbounded medium. The
properties of the medium were expressed by a single dimensionless
parameter $B$ that can be interpreted as a measure of the local excitation
threshold which increases with $B$ while the excitability decreases.
Wave segments exist in a finite range $0\leq B\leq B_{c}$ of $B$-values.

\indent In the limit $B\rightarrow0,$ wave segments correspond to
motionless circular spots \cite{kothe2009second}. At the upper boundary
of the existence range, for $B\rightarrow B_{c}\text{\ensuremath{\approx}}0.535,$
they merge with spiral waves in a separatrix solution known as the
critical finger \cite{karma1991universal}. The critical finger is
an half-infinite plane pulse with an open end that can be regarded
as a spiral wave rotating with zero frequency around an infinitely
large core.

\indent Spiral waves with $B\lesssim B_{c}$ close to the critical
finger and their transition into meandering spiral waves were analytically
investigated by Hakim and Karma \cite{hakim1999theory} applying methods
of singular perturbation theory. For $B>B_{c}$, only retracting fingers
were found to exist because the excitability of the system is too
weak to support wave segments and spiral waves. The entire range $B_{\text{min}}\leq B\leq B_{c}$
for which spiral waves coexist with wave segments as a solution to
the kinematic equations was studied numerically by Zykov in \cite{zykov2007selection,zykov2009kinematics}.
For $B\rightarrow B_{\text{min}}\approx0.211$ the spiral core radius
$r_{0}$ decreases to zero and for $B<B_{\text{min}}$, rigidly rotating
spiral waves cease to exist.

\indent Solving the kinematic equations in a disk with a Neumann
boundary led to the discovery of boundary spots. Boundary spots are
unstable wave solutions to reaction-diffusion equations that rotate
at a lower frequency than spiral waves. Furthermore, in contrast to
spiral waves, boundary spots do not extend over the entire disk but
remain spatially localized close to the disc boundary \cite{bordyugov2007continuation}.

\indent In this work, we follow an analytical non-perturbative approach
that goes back to a classical paper by Burton, Cabrera and Frank \cite{burton1951growth}.
Their approach is non-perturbative in the sense that it does not rely
on a small parameter for a perturbation expansion. These authors considered
spiral waves occurring in crystal growth which have a vanishing core
radius. They used an ansatz function for the wave shape to calculate
the rotation frequency of spirals. Keener \cite{keener1986geometrical}
and Tyson and Keener \cite{tyson1988singular} extended this ansatz
to spirals pinned to a circular hole with finite core radius with
a no-flux boundary condition at the periphery. In this case the selection
problem turns out to be much simpler than for free spiral waves because
the rotation frequency can be determined from the equation for the
front interface alone while the core radius is given by the radius
of the Neumann hole.

\indent Below, we present a non-perturbative approach which does
not only improve the result obtained by Keener and Tyson for pinned
spiral waves, but also works quite well for free spirals. In Sec.
\ref{sec:Equations-and-ansatz}, we state the equations of the free-boundary
approach and review existing solutions. Our ansatz is introduced in
Sec. \ref{sec:Ansatz-and-asymptotic}. The analytical results for
free and pinned spirals are presented in Sec. \ref{sec:Results-for-free}
and Sec. \ref{sec:Results-for-pinned}, respectively, and compared
to numerical solutions of the kinematic equations. We end with discussion
of the results, conclusion and outlook in Sec. \ref{sec:Conclusions}.

\section{\label{sec:Equations-and-ansatz}Kinematic Equations}

\begin{figure}
\includegraphics[scale=0.3]{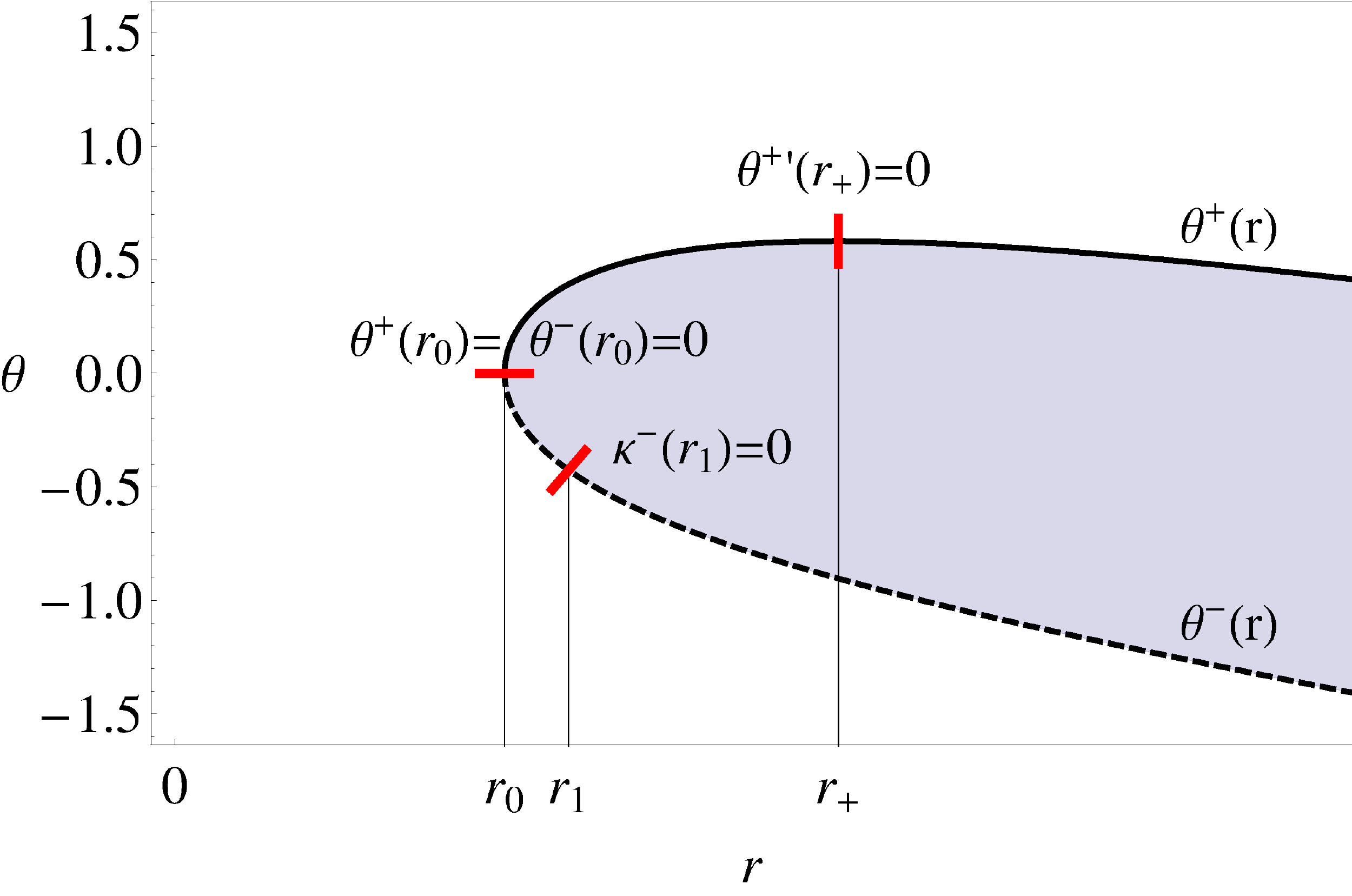} \caption{\label{fig:SpiralWaveTipCloseup}Close-up of the spiral tip region.
Front ($\theta^{+}\left(r\right)$, solid line) and back ($\theta^{-}\left(r\right)$,
dashed line) interface separate the shaded excited region $\mathcal{D}^{+}$
from the white refractory region $\mathcal{D}^{-}$. At the core radius
$r_{0}$, the point $r_{+}$ at the wave front and the inflection
point $r_{1}$ at the wave back, approximate and exact analytical
solution of the free-boundary problem have the same leading order
asymptotic expansions.}
\end{figure}

We consider a standard activator $\left(u\right)$ -inhibitor $\left(v\right)$
reaction-diffusion systems of the form 
\begin{align}
\partial_{t}u & =\epsilon\nabla^{2}u+f\left(u,v\right)/\epsilon,\label{eq:RDSu}\\
\partial_{t}v & =g\left(u,v\right),\label{eq:RDSv}
\end{align}
where the dimensionless parameter $0<\epsilon\ll1$ is a measure for
the time scale separation between activator and inhibitor and serves
as a small parameter for a perturbation expansion. We neglect inhibitor
diffusion and scale space accordingly so that the activator diffusion
coefficient is equal to $\epsilon$. The $u$ nullcline obtained from
$f\left(u,v\right)=0$ is assumed to be $S$-shaped in the $\left(u,v\right)$
plane. A simple choice for the functions $f$ and $g$ is given by
the FitzHugh-Nagumo kinetics 
\begin{align}
f\left(u,v\right) & =3u-u^{3}-v,\label{eq:FHNAcivator}\\
g\left(u,v\right) & =u-\delta,\label{eq:FHNInhibitor}
\end{align}
with a unique, linearly stable rest state $u_{0}=\delta,\, v_{0}=3\delta-\delta^{3}.$\\
If $\epsilon$ is small, a traveling pulse can be regarded as consisting
of two separate spatial regions: an excited region ($\mathcal{D}^{+}$),
where the value of the activator is large and the inhibitor is rising,
and a refractory region ($\mathcal{D}^{-}$), where the activator
value is small and the inhibitor is decaying. This behavior is described
by the outer equations Eqs. \eqref{eq:RDSu}, \eqref{eq:RDSv}, which
in lowest order to $\epsilon$ read \cite{tyson1988singular}
\begin{eqnarray}
0 & = & f\left(u^{\pm}\left(v\right),v\right),\label{eq:FullFreeBoundary1}\\
\partial_{t}v & = & g\left(u^{\pm}\left(v\right),v\right)\;\text{in}\;\mathcal{D}^{\pm}.\label{eq:FullFreeBoundary2}
\end{eqnarray}
Here, $u^{+}\left(v\right)$ and $u^{-}\left(v\right)$ denote the
largest respectively smallest root of $f\left(u,v\right)=0$\emph{
}which the activator follows in the excited respectively refractory
region\emph{. }The two regions $\mathcal{D}^{+}$ and $\mathcal{D}^{-}$
are separated by a front ($+$) and a back ($-$) interface, where
the activator value changes very fast from a low to a high value and
the other way round, respectively. These interfaces can be regarded
as fronts traveling with velocities $c^{\pm}$. They are solutions
to the inner equations, obtained from Eqs. \eqref{eq:RDSu}, \eqref{eq:RDSv}
by a change of scale in time and space proportional to $\epsilon.$
The expression for the front velocity together with Eq. \eqref{eq:FullFreeBoundary2}
and appropriate periodic boundary conditions yield the dispersion
relation for a periodic pulse train, i.e., the dependence of the propagation
velocity $c$ on the period length $L$ to lowest order in $\epsilon$
\cite{keener1986geometrical,tyson1988singular}.\\
In two spatial dimensions, the shape of the front ($+$) and back
($-$) interfaces for rigidly rotating spiral waves are conveniently
parametrized by $\theta^{\pm}\left(r\right)$ using polar coordinates
\begin{align}
\left(\begin{array}{c}
x^{\pm}\left(r,t\right)\\
y^{\pm}\left(r,t\right)
\end{array}\right) & =\left(\begin{array}{c}
r\cos\left(\theta^{\pm}\left(r\right)-\omega t\right)\\
r\sin\left(\theta^{\pm}\left(r\right)-\omega t\right)
\end{array}\right).\label{eq:PolarCoordinates}
\end{align}
In Eq. \eqref{eq:PolarCoordinates}, $\omega>0$ is the rotation frequency
of a spiral wave rotating counterclockwise. The inner equations in
two spatial dimensions provide a relation between the normal velocity
$c_{n}^{\pm}$ of the moving front and back interface and its local
curvature $\kappa^{\pm}$ \cite{keener1986geometrical}, the so-called
linear eikonal equation 
\begin{align}
c_{n}^{\pm}\left(r\right) & =c^{\pm}\left(v^{\pm}\right)-\epsilon\kappa^{\pm}\left(r\right).\label{eq:LinearEikonal}
\end{align}
Here $v^{\pm}$ denote the inhibitor level at the interface, and $c^{\pm}\left(v^{\pm}\right)$
is the velocity of a planar front moving through a medium with a constant
inhibitor value $v^{\pm}$. Similar as for a one-dimensional pulse
train, Eq. \eqref{eq:FullFreeBoundary2} yields together with the
condition of periodicity in $\theta$ an expression for $c^{\pm}\left(v^{\pm}\right).$
This constitutes the so-called wave front interaction model. The interaction
between wave front and wave back is mediated through the dependence
of $v^{+}$ and $v^{-}$ on the positions of both front and back interface.
\\
With the chosen parametrization, the curvature $\kappa^{\pm}$ is
expressed as 
\begin{align}
\kappa^{\pm}\left(r\right) & =-\dfrac{\theta^{\pm}\vspace{0mm}'\left(r\right)}{\left(1+\left(r\theta^{\pm}\vspace{0mm}'\left(r\right)\right)^{2}\right)^{1/2}}-\dfrac{\left(\text{d}/\text{d}r\right)\left(r\theta^{\pm}\vspace{0mm}'\left(r\right)\right)}{\left(1+\left(r\theta^{\pm}\vspace{0mm}'\left(r\right)\right)^{2}\right)^{3/2}},
\end{align}
and the normal velocity is given by 
\begin{align}
c_{n}^{\pm}\left(r\right) & =\dfrac{r\omega}{\left(1+\left(r\theta^{\pm}\vspace{0mm}'\left(r\right)\right)^{2}\right)^{1/2}}.\label{eq:normalvelocity}
\end{align}
Eq. \eqref{eq:LinearEikonal} has to be supplemented with appropriate
boundary conditions. For a rigidly rotating free spiral wave, front
and back interface meet continuously at the apex $r=r_{0}$ of the
spiral, i.e., 
\begin{align}
\theta^{\pm}\left(r_{0}\right) & =0,\label{eq:thetar0}
\end{align}
where we fixed an arbitrary initial phase of the spiral to be zero.
The apex is the point of closest approach of both interfaces to the
center of rotation (compare Fig. \ref{fig:SpiralWaveTipCloseup}).
At the apex, the normal velocity is zero, $c_{n}^{\pm}\left(r_{0}\right)=0.$
Both interfaces approach the apex tangentially to a circle with core
radius $r_{0},$ so that 
\begin{align}
\theta^{+}\vspace{0mm}'\left(r_{0}\right) & =-\theta^{-}\vspace{0mm}'\left(r_{0}\right)=\infty.\label{eq:dthetar0}
\end{align}
This circle is considered as the spiral core with $r_{0}$ being the
core radius. \\
Far from the core, front and back interface behave as an Archimedean
spiral, 
\begin{align}
\theta^{\pm}\left(r\right) & \sim r,\;\left(r\rightarrow\infty\right).\label{eq:thetarinfty}
\end{align}
Eqs. \eqref{eq:thetar0}, \eqref{eq:dthetar0}, \eqref{eq:thetarinfty}
fix six boundary conditions for two coupled second order ordinary
differential equations (ODEs) Eq. \eqref{eq:LinearEikonal}. Four
boundary conditions are necessary to determine the four integration
constants of these ODEs. The remaining two are used to determine two
unknown nonlinear eigenvalues introduced as parameters in the eikonal
equation and the boundary conditions: the rotation frequency $\omega$
and the spiral core radius $r_{0}.$ The full wave front interaction
model, as given by Eq. \eqref{eq:FullFreeBoundary2} together with
the linear eikonal equation Eq. \eqref{eq:LinearEikonal} was solved
numerically by Pelcé and Sun in \cite{pelce1991wave} without any
further approximations for a piecewise linear activator kinetics.
\\
Because Eq. \eqref{eq:FullFreeBoundary2} is too difficult for an
analytical treatment, further approximations are necessary. Assuming
that the inhibitor value $v$ stays always close to the stall level
$v=v_{s}$ given by $c^{\pm}\left(v_{s}\right)=0,$ Eq. \eqref{eq:FullFreeBoundary2}
can be simplified \cite{hakim1999theory} 
\begin{align}
\partial_{t}v & =\dfrac{1}{\tau_{e}}\;\text{in}\;\mathcal{D}^{+},\label{eq:vExcited}\\
\partial_{t}v & =-\dfrac{v-v_{0}}{\tau_{R}}\;\text{in}\;\mathcal{D}^{-},\label{eq:vRefractory}\\
c^{\pm}\left(v^{\pm}\right) & =\alpha\left(v_{s}-v^{\pm}\right),
\end{align}
with the abbreviations 
\begin{align}
\tau_{e} & =\dfrac{1}{g\left(u^{+}\left(v_{s}\right),v_{s}\right)},\\
\tau_{R} & =\dfrac{\partial_{u}f}{\partial_{u}g\partial_{v}f-\partial_{v}g\partial_{u}f}\Bigg|_{u=u^{-}\left(v_{s}\right),\, v=v_{s}}.
\end{align}
This approximation assumes a linear rise of the inhibitor during the
excited period on a time scale 
of the order $\tau_{e},$ followed by an exponential decay during
the refractory period on the time scale $\tau_{R}.$\\
Spiral waves close to the critical finger have a diverging period,
so that the inhibitor value $v^{+}$ has already decayed to its rest
value, $v^{+}=v_{0},$ everywhere along the front interface. In this
case, $v^{-}$ determined by Eqs. \eqref{eq:vExcited}, \eqref{eq:vRefractory}
depends linearly on the angular pulse width $\Delta\theta\left(r\right)=\theta^{+}\left(r\right)-\theta^{-}\left(r\right)$,
and the expressions for $c^{\pm}$ become particularly simple 
\begin{align}
c^{+}\left(v^{+}\right) & =c,\label{eq:cplusfront}\\
c^{-}\left(v^{-}\right) & =-c+\dfrac{b}{\omega}\left(\theta^{+}\left(r\right)-\theta^{-}\left(r\right)\right).\label{eq:cminusback}
\end{align}
$c=\alpha\left(v_{s}-v_{0}\right)>0$ corresponds to the velocity
of a front solution of the inner equations moving through a medium
with a constant inhibitor at its rest state $v=v_{0}.$ Note that
the eikonal equation for the front interface decouples from the equation
for the back, while the back interface interacts with the front interface
via a term that is linear in the pulse width. The single kinetic parameter
$b=\alpha/\tau_{e}>0$ is a measure for the strength of this interaction.
For FitzHugh-Nagumo kinetics according to Eqs. \eqref{eq:FHNAcivator},
\eqref{eq:FHNInhibitor}, we find $\alpha=1/\sqrt{2}$, $v_{s}=0$,
$\tau_{e}=\frac{1}{\sqrt{3}-\delta}$ and $\tau_{R}=6$.\\
Hakim and Karma \cite{hakim1999theory} used singular perturbation
theory to expand the eikonal equation Eq. \eqref{eq:LinearEikonal}
around the critical finger. In that way, they obtain analytical expressions
for spiral waves with a very large core radius. We review their approach
in the following. Taking into account Eqs. \eqref{eq:cplusfront},
\eqref{eq:cminusback}, the eikonal equations for front and back are
\begin{align}
c_{n}^{+}\left(r\right) & =c-\epsilon\kappa^{+}\left(r\right),\label{eq:LinearEikonalFront}\\
c_{n}^{-}\left(r\right) & =\dfrac{b}{\omega}\left(\theta^{+}\left(r\right)-\theta^{-}\left(r\right)\right)-c-\epsilon\kappa^{-}\left(r\right).\label{eq:LinearEikonalBack}
\end{align}
Similar as in the derivation of the eikonal equations from the reaction
diffusion system, $\epsilon$ serves as the small parameter for a
singular perturbation expansion. For both the front and back interface
three scaling regions were identified: the spiral tip region near
to the core, an intermediate region, and one region sufficiently far
from the core where curvature effects are less important. The outer
equations valid in the region far from the core are Eqs. \eqref{eq:LinearEikonalFront},
\eqref{eq:LinearEikonalBack} with $\epsilon=0.$ Its solution 
\begin{align}
\psi_{\text{inv}}^{+}\left(r\right) & =r\theta_{\text{inv}}^{+}\vspace{0mm}'\left(r\right)=-\sqrt{\dfrac{r^{2}\omega^{2}}{c^{2}}-1},\label{eq:InvoluteSpiral}\\
\theta_{\text{inv}}^{-}\left(r\right) & =\theta_{\text{inv}}^{+}\left(r\right)-\dfrac{2c\omega}{b},
\end{align}
describes the involute of a circle of radius $r_{0}$ which asymptotically
transforms into an Archimedean spiral for $r\rightarrow\infty.$\\
The behavior in the tip region is described by the equations for the
critical finger. While the equation for the front can be solved analytically
\cite{zykov2005wave}, no analytical solution is known for the back.
Matching the analytical solutions for the front interface in the tip,
intermediate and far core regions, Hakim and Karma succeeded to derive
analytically a relationship between rotation frequency $\omega$ and
core radius $r_{0}.$ Matching the solutions for the back interface
and using stability arguments, an expression for the core radius $r_{0}$
involving two numerically determined constants was obtained. Together,
these two relations yield the desired dependence of the spiral wave
frequency $\omega$ on the kinetic parameter $b$. It should be emphasized
that these solutions are only valid for small $\epsilon$ and for
spiral waves close to the critical finger which have a diverging core
radius. \\
The eikonal equations for wave front and back, Eqs. \eqref{eq:LinearEikonalFront},
\eqref{eq:LinearEikonalBack}, can be rescaled by introducing dimensionless
quantities according to 
\begin{align}
r_{0} & =\dfrac{R_{0}\epsilon}{c}, & r & =\dfrac{R\epsilon}{c}, & \omega & =\dfrac{c^{2}\Omega}{\epsilon}, & b & =\dfrac{Bc^{3}}{\epsilon}, & c_{n}^{\pm} & =cC_{n}^{\pm},\label{eq:rescaling}
\end{align}
 and rescaled shape functions $\Theta^{\pm}$ as 
\begin{align}
\theta^{\pm}\left(r\right) & =\theta^{\pm}\left(\dfrac{R\epsilon}{c}\right)=\Theta^{\pm}\left(R\right)
\end{align}
 and 
\begin{align}
\Psi^{\pm}\left(R\right) & =R\Theta^{\pm}\vspace{0cm}'\left(R\right).
\end{align}
Here we introduced the dimensionless parameter $B$ as a measure of
the excitation threshold. The rescaling of $b$ by $\epsilon$ is
justified close to the critical finger because there $B\rightarrow B_{c}\text{\ensuremath{\approx}}0.535$
is of order one. In the following all rescaled dimensionless quantities
will be denoted by upper case letters, while lower case letters are
used for dimensional quantities. In dimensionless terms Eq. \eqref{eq:LinearEikonalFront}
and Eq. \eqref{eq:LinearEikonalBack} read 
\begin{align}
C_{n}^{+}\left(R\right) & =1-K^{+}\left(R\right),\label{eq:RescaledEikonalFront}\\
C_{n}^{-}\left(R\right) & =-1+\dfrac{B}{\Omega}\left(\Theta^{+}\left(R\right)-\Theta^{-}\left(R\right)\right)-K^{-}\left(R\right).\label{eq:RescaledEikonalBack}
\end{align}
Note that the small parameter $\epsilon$ as well as the propagation
velocity $c$ in the eikonal equations have been eliminated under
the rescaling.\\
Strictly speaking, these rescaled eikonal equations can only be valid
in the limit of spirals with diverging core radius. The front interface
of spiral waves with finite core radius $R_{0}$ interacts with the
back interface of the wave ahead because it does not propagate into
a fully recovered medium. In general, the front inhibitor level $v^{+}$
depends on the radial coordinate $r$. Zykov \cite{zykov2009kinematics}
introduces a further approximation: assuming a constant value $v^{+}$
of the inhibitor at the front interface, with $v^{+}$ given by the
dispersion relation of a one-dimensional periodic pulse train, and
using a slightly different rescaling, the dimensionless eikonal equations
Eqs. \eqref{eq:RescaledEikonalFront}, \eqref{eq:RescaledEikonalBack}
can be also be used for spirals which are not close to the critical
finger. Applying a numerical shooting method, Zykov \cite{zykov2007selection,zykov2009kinematics}
then proceeds to demonstrate the existence of spiral wave solutions
to these equations in a certain interval $B_{\text{min}}\approx0.211<B\lesssim B_{c}\approx0.535$
of the dimensionless excitability parameter $B$ and determined a
universal relationship $\Omega\left(B\right)$. At $B=B_{\text{min}}\approx0.211$
the shape of the front interface is identical to that obtained by
Burton, Cabrera and Frank (BCF) \cite{burton1951growth} for a spiral
wave with zero core radius rotating at frequency $\Omega\approx0.331$.
The back interface results from turning the front interface clockwise
around an angle $\Delta\Theta\left(R\right)=\Theta^{+}\left(R\right)-\Theta^{-}\left(R\right)=\pi$.
In the other limit, for $B\lesssim B_{c}\approx0.535$, the analytical
results of Hakim and Karma for spirals with diverging core radius
are recovered. The numerically obtained universal relationship $\Omega\left(B\right)$
together with the dispersion relation of one dimensional pulse trains
is sufficient to predict the rotation frequency of rigidly rotating
spiral waves. Though only approximately valid, this approach clearly
separates the two physical mechanisms underlying the frequency selection
for spiral waves:\\
I. The interaction of the front interface with the back interface
of the preceding wave essentially leads to a front moving through
a partially recovered medium. This in turn leads to a slower velocity
of the front as approximately given by the dispersion relation of
a one-dimensional periodic pulse train.\\
II. The interaction of the back interface with the front interface
within the same wave is proportional to the angular pulse width $\Delta\Theta\left(R\right)=\Theta^{+}\left(R\right)-\Theta^{-}\left(R\right)$
and characterized in strength by the dimensionless parameter $B$.\\
That the kinetic characteristics of the medium can be lumped together
into a single parameter \textbf{$B$} simplifies the determination
of the parameter range of spiral wave existence significantly.
\\
\\
\indent Front and back interface of spiral waves pinned to a circular
Neumann hole of radius $r_{+}$ approach the hole in radial direction,
so that $\theta^{\pm}$ fulfills the boundary condition 
\begin{align}
\theta^{\pm}\vspace{0mm}'\left(r_{+}\right) & =0.
\end{align}
which implies that the spiral arm is orthogonal to the hole. Note
that a corresponding point $r=r_{+}$ can always be found at the front
interface of a freely rotating spiral wave, see Fig. \ref{fig:SpiralWaveTipCloseup}.
The kinematic equation for the front interface of a pinned spiral
wave was studied analytically by Keener and Tyson \cite{keener1986geometrical,tyson1988singular}.
These authors determined the asymptotic behavior of solutions to Eq.
\eqref{eq:LinearEikonalFront} for the front interface as $r\rightarrow r_{+}$
and $r\rightarrow\infty$. An ansatz showing the same asymptotic behavior
and involving several constants was used. Comparing the asymptotics
of ansatz and solution, they were able to determine the constants
of the ansatz and finally derived a relation between rotation frequency
$\omega$ and hole radius $r_{+}$.

\section{\label{sec:Ansatz-and-asymptotic}Asymptotes to solutions and ansatz}

\noindent In this section, we propose an attempt to solve the kinematic
Eqs. \eqref{eq:LinearEikonalFront}, \eqref{eq:LinearEikonalBack}
together with the boundary conditions for a free spiral wave Eqs.
\eqref{eq:thetar0}, \eqref{eq:dthetar0}, and \eqref{eq:thetarinfty}.
First, we obtain asymptotes to the solutions to these linear eikonal
equations. Asymptotes to the solutions for the front and back interface
can be obtained at the spiral core, $r\rightarrow r_{0}$, and very
far from the core as $r\rightarrow\infty$. Additionally, an asymptote
can be obtained at the point $r\rightarrow r_{+}$ of the front interface.
Furthermore, the existence of an inflection point at $r=r_{1}$ at
the back interface is taken into account. Second, we present an ansatz
for the interface shape $\theta^{\pm}\left(r\right)$ that reproduces
in leading order all these asymptotic expansions correctly. See e.
g. \cite{bender1978advanced} how to compute asymptotes to solutions
to differential equations.

\subsection{Asymptotes to solutions to the linear eikonal equation}

\noindent Far from the core the shape of the interfaces is asymptotically
Archimedean, i.e., 
\begin{align}
\theta^{\pm}\left(r\right) & =-\dfrac{\omega}{c}r+\mathcal{O}\left(\log\left(r\right)\right),\; r\rightarrow\infty.\label{eq:AsymSol2pm}
\end{align}
At the spiral core, the asymptotic behavior that fulfills the two
boundary conditions Eqs. \eqref{eq:thetar0}, \eqref{eq:dthetar0}
is given by 
\begin{align}
\theta^{\pm}\left(r\right) & =\pm\dfrac{\sqrt{2\epsilon}}{\sqrt{cr_{0}^{2}+\epsilon r_{0}}}\sqrt{r-r_{0}}+\mathcal{O}\left(r-r_{0}\right),\; r\rightarrow r_{0}.\label{eq:AsymSol1pm}
\end{align}
Finally, at the distance $r=r_{+}$ (compare Fig. \ref{fig:SpiralWaveTipCloseup}),
an asymptotic expansion for the front interface is available which
reads \cite{keener1986geometrical,tyson1988singular} 
\begin{align}
\theta^{+}\vspace{0cm}'\left(r\right) & =\dfrac{\left(r_{+}\omega-c\right)}{r_{+}\epsilon}\left(r-r_{+}\right)+\mathcal{O}\left(\left(r-r_{+}\right)^{2}\right),\; r\rightarrow r_{+}.\label{eq:AsymSol3p}
\end{align}
The back interface of spiral waves always exhibits an inflection point
at $r=r_{1}$ where the curvature $\kappa^{-}$ vanishes 
\begin{align}
\kappa^{-}\left(r_{1}\right) & =0.\label{eq:InflectionPoint1}
\end{align}
In polar coordinates the inflection point is not easy visible, see
Fig. \ref{fig:SpiralWaveTipCloseup}, while it appears clearly pronounced
in Cartesian coordinates used in Fig. \ref{fig:FreeSpiralShape}.
From the eikonal equation for the back Eq. \eqref{eq:LinearEikonalBack}
follows at the inflection point 
\begin{align}
c_{n}^{-}\left(r_{1}\right) & =c^{-}\left(r_{1}\right),\label{eq:InflectionPoint2}
\end{align}
which using Eq. \eqref{eq:normalvelocity} leads to the following
expression for the parameter $b$ 
\begin{align}
b & =\frac{\omega}{\theta^{+}\left(r_{1}\right)-\theta^{-}\left(r_{1}\right)}\left(c+\frac{r_{1}\omega}{\sqrt{\left(r_{1}\theta^{-}\vspace{0cm}'\left(r_{1}\right)\right){}^{2}+1}}\right).\label{eq:bExplicit}
\end{align}
Note that although we present here the asymptotes to solutions of
the unscaled eikonal equations, Eqs. \eqref{eq:LinearEikonalFront},
\eqref{eq:LinearEikonalBack}, a rescaling according to Eq. \eqref{eq:rescaling}
yields the corresponding asymptotes of the rescaled eikonal equations
Eqs. \eqref{eq:RescaledEikonalFront}, \eqref{eq:RescaledEikonalBack}.
As it should be the case, $\epsilon$ and $c$ drop out under this
rescaling in every expression for the asymptotes.
\begin{figure}
\includegraphics[scale=0.3]{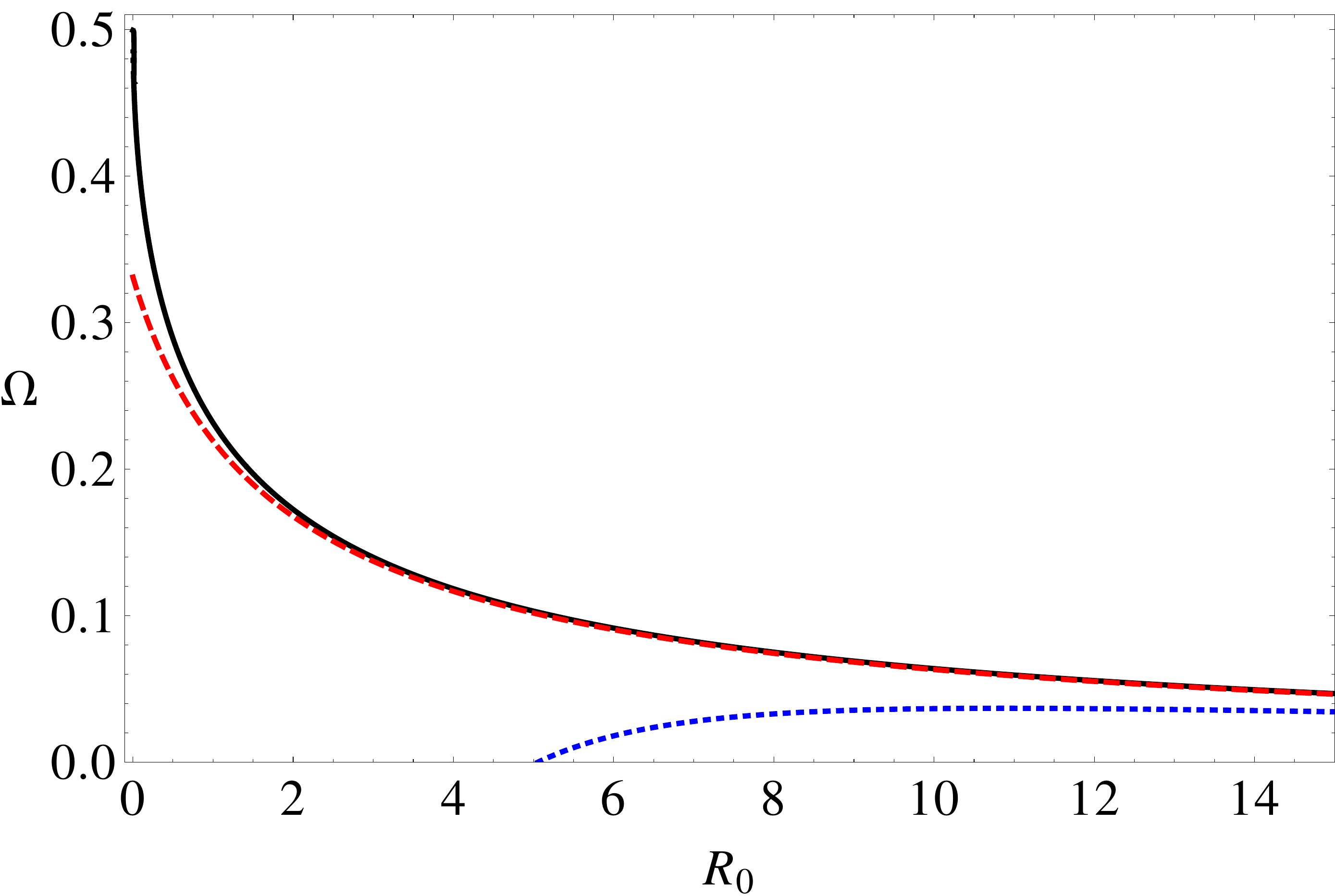}\caption{\label{fig:FigOmegaOverR0}Dimensionless rotation frequency $\Omega$
over dimensionless spiral core radius $R_{0}$ for rigidly rotating
spiral waves. The analytically obtained result according to Eq. \eqref{eq:OmegaOverR0}
(black solid line) is compared to numerical solutions of the linear
eikonal equations Eqs. \eqref{eq:RescaledEikonalFront}, \eqref{eq:RescaledEikonalBack}
(red dashed line). The blue dotted line shows the analytical result
for diverging core radius \cite{hakim1999theory}.}
\end{figure}

\subsection{Asymptotic behavior of the ansatz}

\noindent We use the ansatz 
\begin{align}
\psi_{\text{ans}}^{\pm}\left(r\right) & =r\theta_{\text{ans}}^{\pm}\vspace{0mm}'\left(r\right)=A_{\pm}\dfrac{r^{2}-r_{\pm}^{2}}{\sqrt{r^{2}-r_{0}^{2}}},\label{eq:AnsatzPsi}
\end{align}
which yields for the interface shape $\theta_{\text{ans}}^{\pm}\left(r\right)$
\begin{align}
\theta_{\text{ans}}^{\pm}\left(r\right) & =\intop_{r_{0}}^{r}\text{d}\tilde{r}\dfrac{\psi_{\text{ans}}^{\pm}\left(\tilde{r}\right)}{\tilde{r}}\nonumber \\
 & =A_{\pm}\left(\sqrt{r^{2}-r_{0}^{2}}-\dfrac{r_{\pm}^{2}}{r_{0}}\arccos\left(\dfrac{r_{0}}{r}\right)\right).\label{eq:AnsatzTheta}
\end{align}
Such an ansatz can only be justified by virtue of the validity of
the conclusions derived from it. It involves five constants $A_{\pm}$,
$r_{\pm}$ and $r_{0}$. For $r_{-}\leq r_{0},$ the back interface
described by $\theta_{\text{ans}}^{-}$ exhibits an inflection point
at a point $r_{1}\geq r_{0}$. Thus, valid solutions for the front
and back of a free spiral can only be found if 
\begin{align}
r_{+} & \geq r_{0}\geq r_{-}\ge0.\label{eq:Conditionrpm}
\end{align}
So in contrast to the spiral core radius $r_{0}$ and the corresponding
Neumann hole radius $r_{+}$, $r_{-}$ does not have a direct physical
interpretation. All five constants together with the spiral wave frequency
$\omega$ are determined by comparing the asymptotics of the ansatz
with the asymptotes to the solution to the eikonal equations. Our
ansatz produces the correct leading order asymptotics for $r\text{\ensuremath{\rightarrow\infty}}$
\begin{align}
\theta_{\text{ans}}^{\pm}\left(r\right) & =A_{\pm}r+\mathcal{O}\left(1\right),\; r\rightarrow\infty,\label{eq:AsymAns1pm}
\end{align}
and for $r\rightarrow r_{0}$ 
\begin{align}
\theta_{\text{ans}}^{\pm}\left(r\right) & =A_{\pm}\dfrac{\sqrt{2}}{r_{0}^{3/2}}\left(r_{0}^{2}-r_{\pm}^{2}\right)\sqrt{r-r_{0}}\nonumber \\
 & +\mathcal{O}\left(\left(r-r_{0}\right)^{3/2}\right),\; r\rightarrow r_{0}.\label{eq:AsymAns2pm}
\end{align}
At the point $r=r_{+},$ the ansatz for the front interface displays
the asymptotic behavior corresponding to a Neumann boundary 
\begin{align}
\theta_{\text{ans}}^{+}\vspace{0cm}'\left(r\right) & =\dfrac{2A_{+}}{\sqrt{r_{+}^{2}-r_{0}^{2}}}\left(r-r_{+}\right)+\mathcal{O}\left(\left(r-r_{+}\right)^{2}\right),\; r\rightarrow r_{+}.\label{eq:AsymAnsrp}
\end{align}
The sixth relation is the inflection point at $r=r_{1}$ with a vanishing
curvature at the spiral wave back, Eq. \eqref{eq:InflectionPoint1}.
Equating the six asymptotic expansions of the ansatz with the six
asymptotes to the solutions to the linear eikonal equation, we are
able to determine four unknown parameters $A_{\pm}$ and $r_{\pm},$
as well as the relations $\omega$ over $r_{0}$ and $r_{0}$ over
$b.$

\section{\label{sec:Results-for-free}Results for free spirals}

\noindent 
\begin{figure}
\includegraphics[scale=0.325]{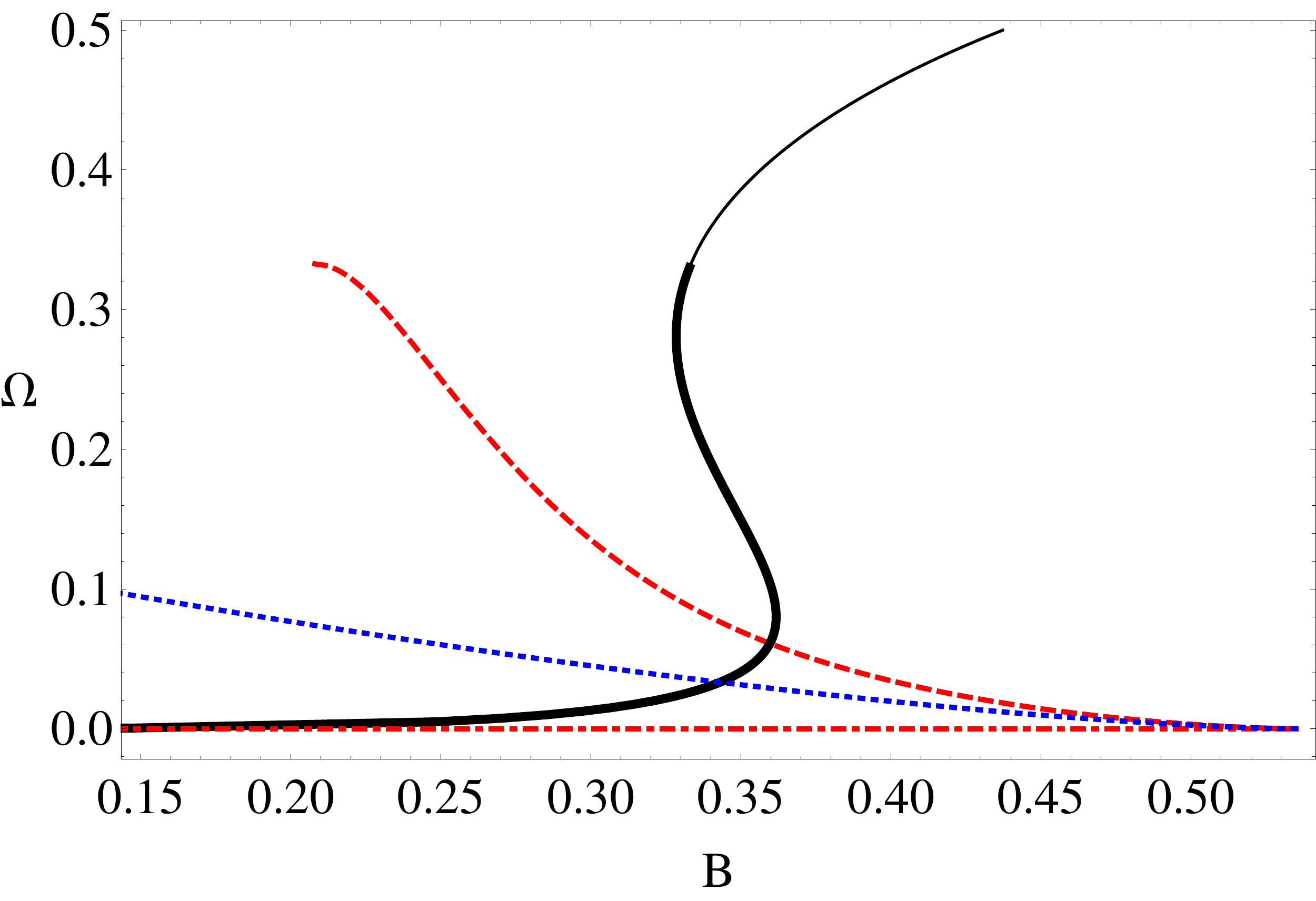} \caption{\label{fig:OmegaOverB}Rotation frequency $\Omega$ versus excitability
parameter $B$ in dimensionless units. The red dashed and dash-dotted
lines represent the branches of spiral waves respectively wave segments
obtained numerically from the linear eikonal equations Eqs. \eqref{eq:RescaledEikonalFront},
\eqref{eq:RescaledEikonalBack}. The two branches merge for $B=B_{c}\approx0.535$
(critical finger \cite{karma1991universal}). The analytical approximation
is shown by the black solid line where the thick segment corresponds
to the frequencies below $\Omega=\Omega_{\text{BCF}}\approx0.331$
(Burton-Cabrera-Frank limit). The blue dotted line shows the analytical
result from \cite{hakim1999theory}.}
\end{figure}

\subsection{Rotation frequency $\Omega$ versus core radius $R_{0}$}

Comparing Eq. \eqref{eq:AsymSol1pm} with Eq. \eqref{eq:AsymAns1pm},
and Eq. \eqref{eq:AsymSol2pm} with Eq. \eqref{eq:AsymAns2pm}, we
get 
\begin{align}
A_{\pm} & =-\dfrac{\omega}{c},\\
r_{\pm} & =\sqrt{r_{0}\left(r_{0}\pm\dfrac{c}{\omega}\sqrt{\frac{\epsilon}{cr_{0}+\epsilon}}\right)}.\label{eq:rpm}
\end{align}
Equating the asymptotic expressions for $r\rightarrow r_{+}$ given
by Eqs. \eqref{eq:AsymAnsrp} and \eqref{eq:AsymSol3p} we obtain
an implicit relation between frequency $\omega$ and core radius $r_{0}$
\begin{align}
\omega & =\sqrt{\frac{c^{2}\omega}{cr_{0}\sqrt{\frac{\epsilon}{cr_{0}+\epsilon}}+r_{0}^{2}\omega}}-2\sqrt{\frac{\omega^{3}\sqrt{\epsilon^{3}\left(cr_{0}+\epsilon\right)}}{c^{3}r_{0}}}.\label{eq:omegaoverr0}
\end{align}
In rescaled form, Eq. \eqref{eq:omegaoverr0} reads 
\begin{align}
\sqrt{R_{0}}\Omega & =\sqrt{\frac{\sqrt{R_{0}+1}\Omega}{1+R_{0}\sqrt{R_{0}+1}\Omega}}-2\sqrt{\Omega^{3}\sqrt{R_{0}+1}}.\label{eq:OmegaOverR0}
\end{align}
The last expression can be written as 
\begin{align}
\tilde{\Omega}\left(1+\tilde{\Omega}\right)\left(R_{0}+2\sqrt{\tilde{\Omega}}\right)^{2}-R_{0}^{2}\left(1+R_{0}\right) & =0\label{eq:OmegaTildeOverR0}
\end{align}
where we have we have introduced the abbreviation 
\begin{align}
\tilde{\Omega} & =R_{0}\sqrt{1+R_{0}}\Omega
\end{align}
which can be used to rewrite $R_{\pm}$ from Eq. \eqref{eq:rpm} in
the form 
\begin{align}
R_{\pm} & =\sqrt{R_{0}\left(R_{0}\pm\dfrac{1}{\sqrt{1+R_{0}}\Omega}\right)}=R_{0}\sqrt{1\pm\dfrac{1}{\tilde{\Omega}}}.\label{eq:Rpm}
\end{align}

\noindent From Descartes rule of signs we conclude that the number
of positive real roots of the sextic polynomial Eq. \eqref{eq:OmegaTildeOverR0}
is one. Thus, $\Omega$ over $R_{0}$ has only one physically meaningful
branch.

\noindent \indent Unfortunately, it is not possible to determine
an explicit relation for $\Omega\left(R_{0}\right)$ from Eq. \eqref{eq:OmegaOverR0}
or Eq. \eqref{eq:OmegaTildeOverR0}. However, we can determine the
asymptotic behavior of $\Omega\left(R_{0}\right)$ for large and small
core radii $R_{0}$ as
\begin{align}
\Omega & =\dfrac{1}{2}-\dfrac{1}{2\sqrt{2}}\sqrt{R_{0}}+o\left(\sqrt{R_{0}}\right),\; R_{0}\rightarrow0
\end{align}
and 
\begin{align}
\Omega & =\dfrac{1}{R_{0}}-\dfrac{1}{2R_{0}^{3/2}}+\mathcal{O}\left(R_{0}^{-7/4}\right),\; R_{0}\rightarrow\infty.
\end{align}
Thus, we obtain a finite rotation frequency for vanishing core radius.
However, with $\Omega\left(R_{0}\rightarrow0\right)=1/2$ we miss
the Burton-Cabrera-Frank limit $\Omega\approx0.331$.\\
Our theoretical prediction for $\Omega\left(R_{0}\right)$ matches
well with direct numerical solutions of the rescaled eikonal equations
Eqs. \eqref{eq:RescaledEikonalFront}, \eqref{eq:RescaledEikonalBack}
(compare black solid respectively red dashed line Fig. \ref{fig:FigOmegaOverR0})
where in particular the good agreement for intermediate and even quite
small core radii is remarkable.

\noindent \indent Hakim and Karma found with singular perturbation
theory for the case of very large core radius $r_{0}$ and small $\epsilon$
\cite{hakim1999theory} 
\begin{align}
\omega_{\text{HK}}\left(r_{0}\right) & =\dfrac{c}{r_{0}}+2^{1/3}\dfrac{a_{1}}{r_{0}^{5/3}}\epsilon^{2/3},\label{eq:omegaKarma}
\end{align}
which in rescaled form reads 
\begin{align}
\Omega_{\text{HK}}\left(R_{0}\right) & =\dfrac{1}{R_{0}}+2^{1/3}\dfrac{a_{1}}{R_{0}^{5/3}},\label{eq:omegaKarmaRescaled}
\end{align}
where $a_{1}=-2.3381$ denotes the first zero of the Airy function
$\text{Ai}\left(x\right).$ The dependence Eq. \eqref{eq:omegaKarmaRescaled}
corresponds to the blue dotted line in Fig. \ref{fig:FigOmegaOverR0}.
To compare our result with the result by Hakim and Karma, we can expand
our result Eq. \eqref{eq:omegaoverr0} for small $\epsilon$, 
\begin{align}
\omega\left(r_{0}\right) & =\dfrac{c}{r_{0}}-\dfrac{c^{1/2}}{2r_{0}^{3/2}}\epsilon^{1/2}+\mathcal{O}\left(\epsilon^{3/4}\right).\label{eq:omegaoverr0perturbation}
\end{align}
Note the different exponents in $\epsilon$ and $r_{0}$.

\noindent \indent Another justification of our ansatz is given by
the following observation. For $\epsilon=0,$ Eq. \eqref{eq:rpm}
and \eqref{eq:omegaoverr0} reduce to 
\begin{align}
\omega & =\dfrac{c}{r_{0}},\\
r_{+} & =r_{0},
\end{align}
 i.e., our ansatz reduces to the involute spiral 
\begin{align}
\lim_{\epsilon\rightarrow0}\psi_{\text{ans}}^{+}\left(r\right) & =-\sqrt{\dfrac{r^{2}}{r_{0}^{2}}-1}.\label{eq:Involut}
\end{align}
Eq. \eqref{eq:Involut} gives the correct solution of the linear eikonal
equation for the front interface for $\epsilon=0$, compare Eq. \eqref{eq:InvoluteSpiral}.

\subsection{\label{sub:OmegaOverB}Rotation frequency $\Omega$ as a function
of $B$}

\noindent 
\begin{figure}
\includegraphics[scale=0.3]{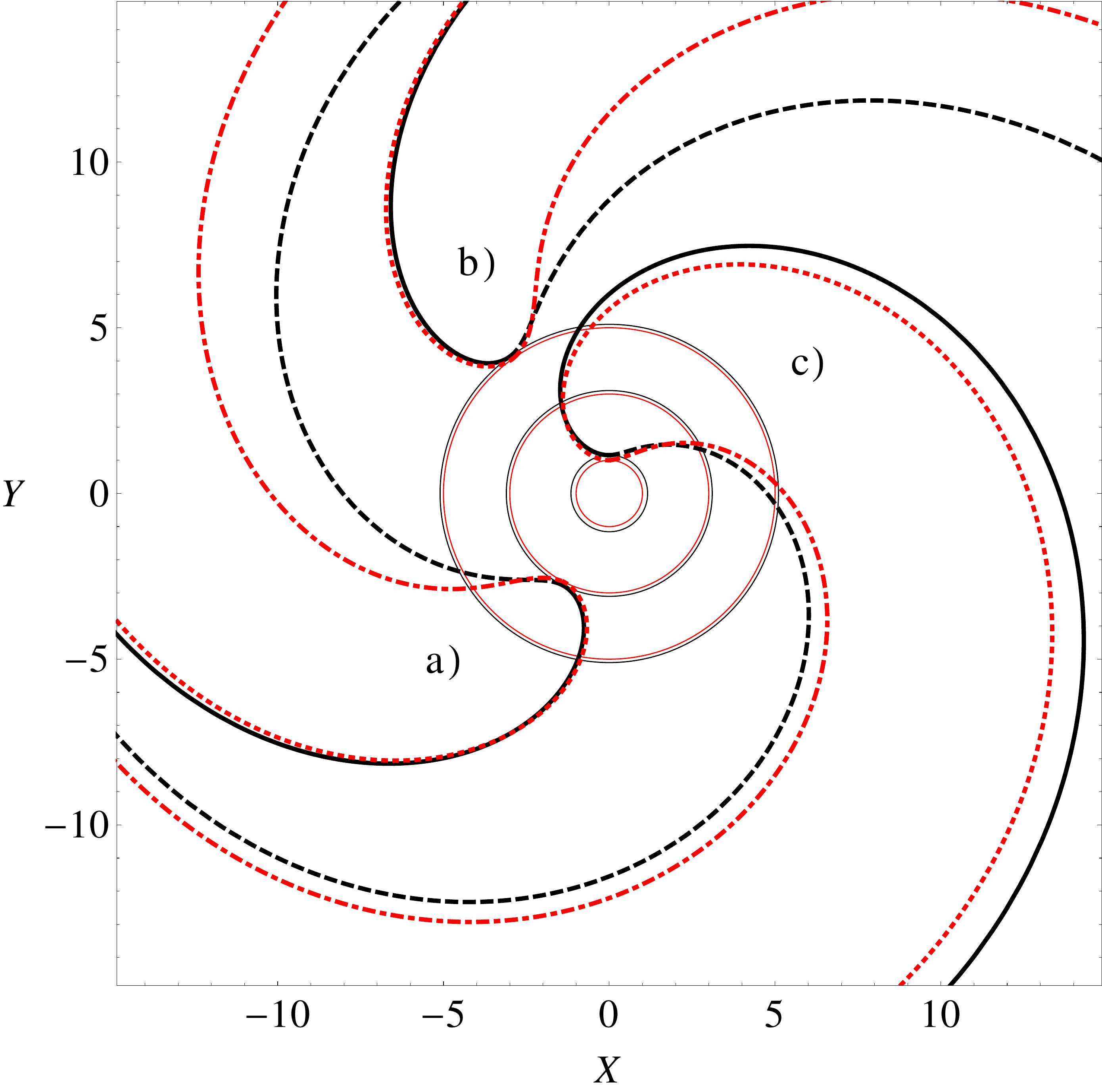} \caption{\label{fig:FreeSpiralShape}Front and back interface of free spiral
waves in the core region. Black solid and dashed lines show analytical
approximations for wave front respectively back as obtained from Eq.
\eqref{eq:FreeSpiralShape} using $\Omega\left(R_{0}\right)$ given
by Eq. \eqref{eq:OmegaOverR0}. Red dotted and dot-dashed lines are
plots of the corresponding numerical results. The rotation frequency
was fixed to a) $\Omega=0.137$, b) $\Omega=0.102$, c) $\Omega=0.219$.
The analytically obtained core radii (black circles) are a) $R_{0}=3$,
b) $R_{0}=5$, c) $R_{0}=1.$}
\end{figure}
Upon rescaling \eqref{eq:bExplicit} and comparing it with the ansatz,
we can express $B$ in the form 
\begin{align}
B & =\frac{\Omega+\frac{R_{1}\Omega^{2}}{\sqrt{R_{1}^{2}\left(\Theta_{\text{ans}}^{-}\vspace{0cm}'\left(R_{1}\right)\right){}^{2}+1}}}{\Theta_{\text{ans}}^{+}\left(R_{1}\right)-\Theta_{\text{ans}}^{-}\left(R_{1}\right)}.\label{eq:BExplicit}
\end{align}
Here, $R_{1}$ is given by 
\begin{align}
K^{-}\left(R_{1}\right) & =0
\end{align}
or explicitly 
\begin{align}
R_{1}^{2}\Theta_{\text{ans}}^{-}\vspace{0cm}'\left(R_{1}\right)^{3}+R_{1}\Theta_{\text{ans}}^{-}\vspace{0cm}''\left(R_{1}\right)+2\Theta_{\text{ans}}^{-}\vspace{0cm}'\left(R_{1}\right) & =0.\label{eq:InflectionPointAnsatz}
\end{align}
If we numerically solve the algebraic Eqs. \eqref{eq:BExplicit} and
\eqref{eq:InflectionPointAnsatz} together with the ansatz Eq. \eqref{eq:AnsatzTheta}
and use the relation for $\Omega$ over $R_{0}$ Eq. \eqref{eq:OmegaOverR0},
we obtain the dependence $\Omega\left(B\right)$ plotted as black
solid line in Fig. \ref{fig:OmegaOverB}. Analytically we can determine
$B$ as a function of $\Omega$ in the limit $\Omega\rightarrow\frac{1}{2}$
corresponding to $R_{0}\rightarrow0$ 
\begin{align}
\lim_{\Omega\rightarrow\frac{1}{2}}B\left(\Omega\right) & =\dfrac{1}{4\text{arcsec}\left(2^{1/4}\right)}\approx0.437171,
\end{align}
and in the limit $\Omega\rightarrow0$ corresponding to $R_{0}\rightarrow\infty$
\begin{align}
\lim_{\Omega\rightarrow0}B\left(\Omega\right)= & 0.
\end{align}
From the obtained $\Omega\left(B\right)$ dependence we conclude that
spiral wave solutions to the linear eikonal equation exist within
this finite range of $B$ values. However, our ansatz clearly fails
to give a satisfying solution for $\Omega\left(B\right)$ upon comparison
with a numerically obtained solution. The red dashed line in Fig.
\ref{fig:OmegaOverB} shows $\Omega\left(B\right)$ calculated numerically
for free spiral waves in \cite{zykov2007selection}. The red dot-dashed
line in Fig. \ref{fig:OmegaOverB} we have added for completeness.
It describes the branch of unstable wave segments. At $B=B_{c}\approx0.535$
the two red branches merge in the critical finger. The blue dotted
line in Fig. \ref{fig:OmegaOverB} shows the analytical result derived
by Hakim and Karma \cite{hakim1999theory} 
\begin{align}
\Omega_{\text{HK}}\left(B\right) & =\dfrac{1}{\sqrt{2}}\left(\dfrac{B-B_{c}}{Ka_{1}}\right)^{3/2},\label{eq:OmegaBKarma}
\end{align}
which is valid close to the critical finger $B\lesssim B_{c}$. In
Eq. \eqref{eq:AnsatzTheta} $K\approx0.630$ and $B_{c}$ are numerically
determined constants.\\
Although our relation for $\Omega\left(B\right)$ is very inaccurate,
it still bears some some resemblance to real spirals in a certain
range of frequencies. In the following we try to analyze the reasons
for its failure. First of all, the analytical solution for $\Omega\left(R_{0}\right)$
for the front interface yields a range of frequencies $0.331\lesssim\Omega\leq1/2$
which cannot be found in numerical solutions. This range corresponds
to the thin solid line in Fig. \ref{fig:OmegaOverB}. The appearance
of that branch is therefore due to the failure of the ansatz for the
front interface in this range of frequencies. Second, the lower branch
resembles wave segments which can be seen as solutions which rotate
with zero frequency. However, our ansatz fails to describe solutions
with small rotation frequency. The shape of the wave front $\Theta^{+}\left(R\right)$
obtained from the rescaled linear eikonal equation Eq. \eqref{eq:RescaledEikonalFront}
behaves in the limit $R\rightarrow\infty$ as 
\begin{align}
\Theta^{+}\left(R\right) & =-\Omega R-\Omega\ln\left(R\right)\nonumber \\
 & +\left(\Omega-\dfrac{1}{2\Omega}\right)\dfrac{1}{R}+\mathcal{O}\left(\dfrac{1}{R^{2}}\right),\; R\rightarrow\infty.\label{eq:ThetaPHigherOrderAsymptotics}
\end{align}
As $B\rightarrow B_{c}$ and $\Omega\rightarrow0$, all terms linear
in $\Omega$ vanish while the term $\sim1/\Omega$ grows indefinitely.
Therefore, expansion Eq. \eqref{eq:ThetaPHigherOrderAsymptotics}
breaks down close to the critical finger. The correct leading order
asymptotics for the critical finger reads 
\begin{align}
\Theta^{+}\left(R\right) & =2\dfrac{\ln\left(R\right)}{R}+o\left(\dfrac{\ln\left(R\right)}{R}\right),\; R\rightarrow\infty,\, B=B_{c}.
\end{align}
The reason for the breakdown of the expansion Eq. \eqref{eq:ThetaPHigherOrderAsymptotics}
for $\Omega\rightarrow0$ is that polar coordinates are a convenient
parametrization for spiral waves but a bad choice for wave segments
and critical finger which are better parametrized in Cartesian coordinates.
Note, that the asymptotics for $r\rightarrow r_{0}$, Eq. \eqref{eq:AsymSol1pm},
remains valid for the critical finger and wave segments. In other
words, the tip region of the critical finger and of wave segments
is correctly represented by our ansatz, however, it fails in correctly
predicting the whole shape of the front interface for wave segments
and the critical finger.

\subsection{\label{sub:AnalyticalApproximationForTheSpiralShape}Analytical approximation
for the spiral shape}

\begin{figure}
\includegraphics[scale=0.3]{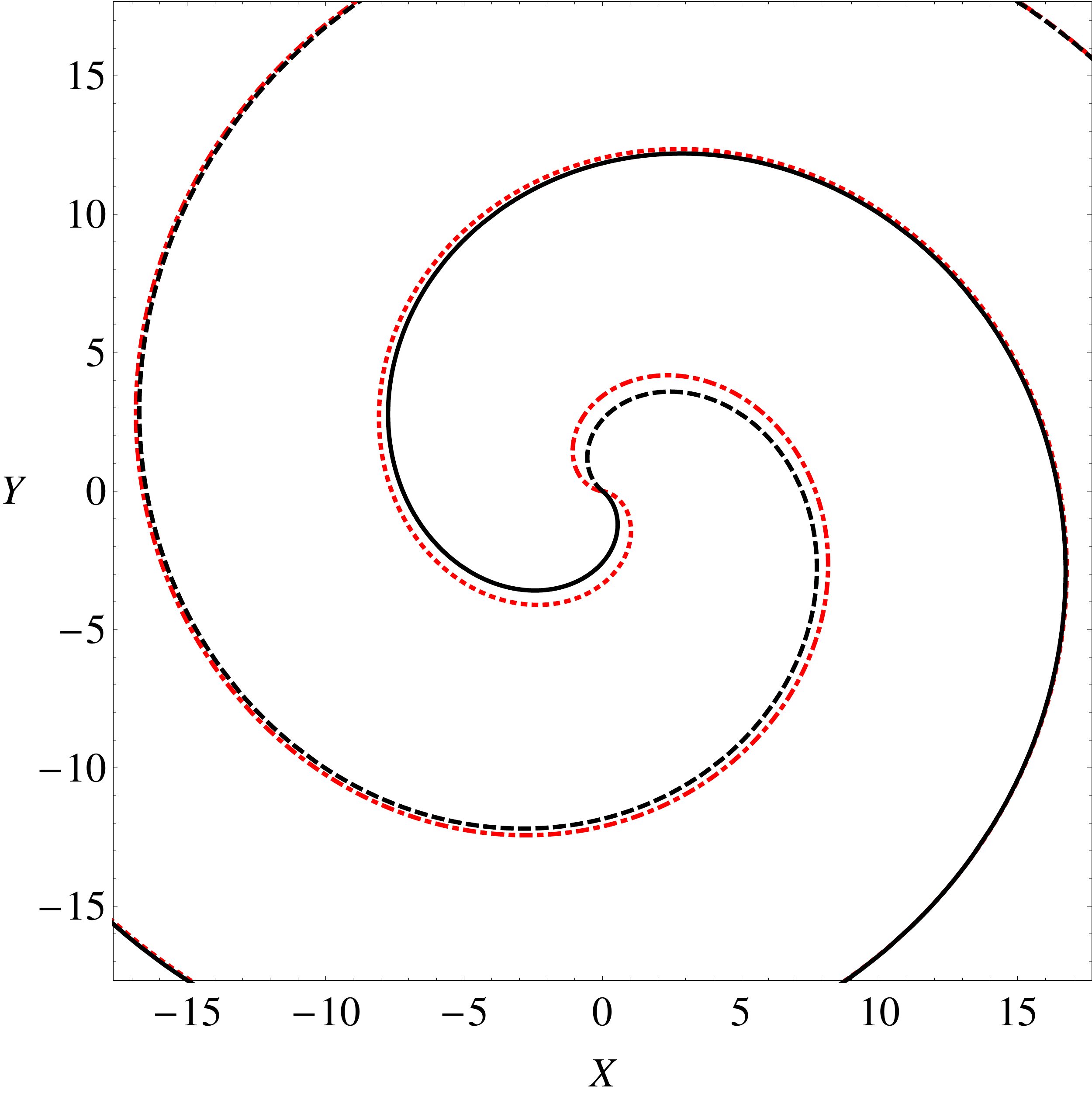}

\caption{\label{fig:BCFSymmetricSpiral}Spiral shape in the BCF limit of zero
spiral core radius. The back interface is identical to the front interface
turned by an angle $\Delta\Theta=\pi$. Numerical solution of the
eikonal equation (red) and Archimedean approximation given by Eq.
\eqref{eq:ArchimedeanApproximation} (black) are plotted in such a
way that they coincide far from the core. This leads to small deviations
near to the center of rotation.}
\end{figure}

\noindent Our ansatz leads to the following analytical approximation
for the front and back interface of a rigidly rotating spiral wave
\begin{align}
\Theta_{\text{ans}}^{\pm}\left(R\right) & =-\sqrt{R^{2}-R_{0}^{2}}\Omega\nonumber \\
 & +\left(R_{0}\Omega\pm\dfrac{1}{\sqrt{R_{0}+1}}\right)\arccos\left(\dfrac{R_{0}}{R}\right),\label{eq:FreeSpiralShape}
\end{align}
where $\Omega$ as a function of core radius $R_{0}$ is given by
Eq. \eqref{eq:OmegaOverR0}. In Fig. \ref{fig:FreeSpiralShape} we
compare the analytical prediction to numerical solutions of the dimensionless
eikonal equations Eqs. \eqref{eq:RescaledEikonalFront}, \eqref{eq:RescaledEikonalBack}
for three given values of the rotation frequency $\Omega$ in order
to avoid the inaccuracy in the analytical relation $\Omega\left(B\right)$.
Fig. \ref{fig:FreeSpiralShape} shows good agreement between theoretical
and numerical results. In particular, the front interface is nicely
described by the ansatz although the analytical approximation always
slightly overestimates the core size. The back interface is well represented
for a small core radius $R_{0}$ but the agreement becomes worse for
larger core radii. The reason is that the analytically predicted pulse
width 
\begin{align}
\Delta\Theta_{\text{ans}}\left(R\right) & =2\dfrac{\arccos\left(\dfrac{R_{0}}{R}\right)}{\sqrt{1+R_{0}}}
\end{align}
displays deviations which increase for large core radii. In fact,
asymptotically we find for large $R$ 
\begin{align}
\Delta\Theta_{\text{ans}}\left(R\right) & =\dfrac{\pi}{\sqrt{1+R_{0}}}+\mathcal{O}\left(\dfrac{1}{R}\right),\; R\rightarrow\infty,\label{eq:OmegaR_0_InfEikonal}
\end{align}
while the asymptotic behavior to the eikonal equation yields 
\begin{align}
\Delta\Theta\left(R\right) & =\Theta^{+}\left(R\right)-\Theta^{-}\left(R\right)\nonumber \\
 & =\dfrac{2\Omega}{B}+\mathcal{O}\left(\dfrac{1}{R}\right),\; R\rightarrow\infty.\label{eq:OmegaR_0_Inf}
\end{align}
If we plug the asymptotic expansion for the rotation frequency 
\begin{align}
\Omega & =\dfrac{1}{R_{0}}+o\left(\dfrac{1}{R_{0}}\right),\; R_{0}\rightarrow\infty,
\end{align}
into \eqref{eq:OmegaR_0_Inf} we get a different asymptotic behavior
for large core radii as compared to \eqref{eq:OmegaR_0_InfEikonal}.
This difference explains the decreasing agreement between analytically
and numerically calculated back interface with increasing $R_{0}$.
Moreover, because $B$ is intimately connected with the pulse width,
it is another reason for the failure of the analytical $\Omega$ over
$B$ relation in the vicinity of the critical finger.\\
The opposite limit, $R_{0}\rightarrow0$, implies $\Omega\rightarrow1/2$,
$B\rightarrow1/\left(4\text{arcsec}\left(2^{1/4}\right)\right)\approx0.437171$
for the ansatz solution. This yields an Archimedean spiral 
\begin{align}
\lim_{R_{0}\rightarrow0}\Theta_{\text{ans}}^{\pm}\left(R\right) & =-\dfrac{1}{2}\left(R\mp\pi\right),
\end{align}
with a back interface identical in shape to the front interface but
turned by an angle $\Delta\Theta_{\text{ans}}\left(R\right)=\pi$.
Numerically solving the kinematic equations for zero core radius,
i.e., for $\Omega\approx0.331$ and $B=B_{\text{min}}\approx0.211$,
leads to a very similar result, compare red lines in Fig. \ref{fig:BCFSymmetricSpiral}
and \cite{zykov2009kinematics,Zykov2011Selection}.\\
To find even better analytical estimates for $B$ and $\Omega$ in
the limit $R_{0}\rightarrow0$, we choose an Archimedean spiral with
the correct leading order asymptotics as $R\rightarrow\infty$, i.e.
\begin{align}
\Theta_{\text{ans}}^{+}\left(R\right) & =-\Omega R,\label{eq:ArchimedeanApproximation}
\end{align}
for the front interface, and assume an identical back interface turned
by an angle $\Delta\Theta_{\text{ans}}\left(R\right)=\pi=\dfrac{2\Omega}{B}$
\cite{Zykov2011Selection}. With this ansatz, we minimize the functional
\begin{align}
S^{+} & =\intop_{0}^{\infty}dr\left(1-K^{+}\left(r\right)-C_{n}^{+}\left(r\right)\right)^{2}\nonumber \\
 & =\dfrac{19}{16}\pi\Omega+\dfrac{4-\pi}{2\Omega}-\log\left(4\right)-1
\end{align}
with respect to $\Omega$. We obtain 
\begin{align}
S^{+} & =\sqrt{\dfrac{19}{8}\left(4-\pi\right)\pi}-1-\log\left(4\right)\approx0.144
\end{align}
for 
\begin{align}
\Omega & =\sqrt{\frac{8\left(4-\pi\right)}{19\pi}}\approx0.339.
\end{align}
The corresponding value for $B$ is 
\begin{align}
B & =\dfrac{2\Omega}{\pi}=4\sqrt{\dfrac{2\left(4-\pi\right)}{19\pi^{3}}}\approx0.216.
\end{align}
These values, though only approximately valid, display a relative
error of less than $3\%$ when compared with the numerical results
of the BCF limit. We compare this Archimedean approximation for the
spiral shape with numerical solutions of the rescaled eikonal equations
Eqs. \eqref{eq:RescaledEikonalFront}, \eqref{eq:RescaledEikonalBack}
in Fig. \ref{fig:BCFSymmetricSpiral}.

\section{\label{sec:Results-for-pinned}Results for pinned spirals}

\noindent In this section we consider a pinned spiral, i. e. a wave
rotating around a hole of radius $r=r_{+}$ with no flux boundaries
at the hole boundary. We are looking for solutions of Eqs. \eqref{eq:LinearEikonalFront},
\eqref{eq:LinearEikonalBack} subject to the following boundary conditions
\begin{align}
\theta^{+}\left(r_{+}\right) & =\theta_{0}^{+},\\
\theta^{\pm}\vspace{0mm}'\left(r_{+}\right) & =0,\\
\theta^{\pm}\left(r\right) & \sim r,\; r\rightarrow\infty.
\end{align}
Four out of these five conditions are necessary to determine four
integration constants of the ODEs. The fifth condition yields a relation
between frequency $\omega$ and hole radius $r_{+}.$\\
The front interface of a free spiral always contains a corresponding
pinned spiral because the front interface displays the point $r=r_{+}$
with $\theta^{\pm}\vspace{0mm}'\left(r_{+}\right)=0$. Therefore,
the same ansatz Eq. \eqref{eq:AnsatzTheta} as was used for free spirals
can be used for the front interface of a pinned spiral. Its back interface
is identical in shape to the front interface but rotated by an angular
puls width $\Delta\Theta$.

\subsection{Rotation frequency $\Omega$ versus hole radius $R_{+}$}

Eliminating $R_{0}$ from the $\Omega\left(R_{0}\right)$ relation
for free spirals, Eq. \eqref{eq:OmegaOverR0} by using Eq. \eqref{eq:Rpm},
we obtain 
\begin{align}
R_{+}\Omega & =R_{+}\Omega\left(1-\frac{4\Omega^{2}}{\left(R_{+}\Omega-1\right){}^{2}}\right)\nonumber \\
 & +\frac{\sqrt{1-\frac{4\Omega^{2}}{\left(R_{+}\Omega-1\right){}^{2}}}}{\sqrt{\sqrt{R_{+}^{2}-\frac{4\Omega^{2}R_{+}^{2}}{\left(R_{+}\Omega-1\right){}^{2}}}+1}}.\label{eq:OmegaOverR+}
\end{align}
Because $R_{+}$ is the given hole radius, \eqref{eq:OmegaOverR+}
already represents the desired result. Note that in contrast to free
spiral waves, for a pinned spiral wave the rotation frequency does
not depend on the parameter $B$ that characterizes the strength of
the interaction between wave front and back and measures the excitability
of the medium. Within our approach, any dependence of the rotation
frequency on $B$ or other kinetic parameters enters through the dispersion
relation for periodic pulse trains.\\
As before, for small $\epsilon$ an explicit expression can be derived
perturbatively 
\begin{align}
\omega\left(r_{+}\right) & =\dfrac{c}{r_{+}}-\dfrac{2c^{1/4}}{r_{+}^{7/4}}\epsilon^{3/4}+\mathcal{O}\left(\epsilon^{5/4}\right).\label{eq:OmegaSmallEps}
\end{align}
Eq. \eqref{eq:OmegaSmallEps} can be compared with Keener's result
\cite{tyson1988singular} 
\begin{align}
\omega_{\text{K}}\left(r_{+}\right) & =\dfrac{c\left(4cr_{+}+\epsilon-\sqrt{\epsilon\left(8cr_{+}+\epsilon\right)}\right)}{4r_{+}\left(cr_{+}+\epsilon\right)},
\end{align}
which after expanding for small $\epsilon$ gives 
\begin{align}
\omega_{\text{K}}\left(r_{+}\right) & =\dfrac{c}{r_{+}}-\dfrac{c^{1/2}}{\sqrt{2}r_{+}^{3/2}}\epsilon^{1/2}+\mathcal{O}\left(\epsilon\right).\label{eq:OmegaSmallEpsKeener}
\end{align}
In rescaled form, Keener's result reads 
\begin{align}
\Omega_{\text{K}}\left(R_{+}\right) & =\dfrac{1+4R_{+}-\sqrt{1+8R_{+}}}{4R_{+}\left(1+R_{+}\right)}.
\end{align}
 
\begin{figure}
\includegraphics[scale=0.3]{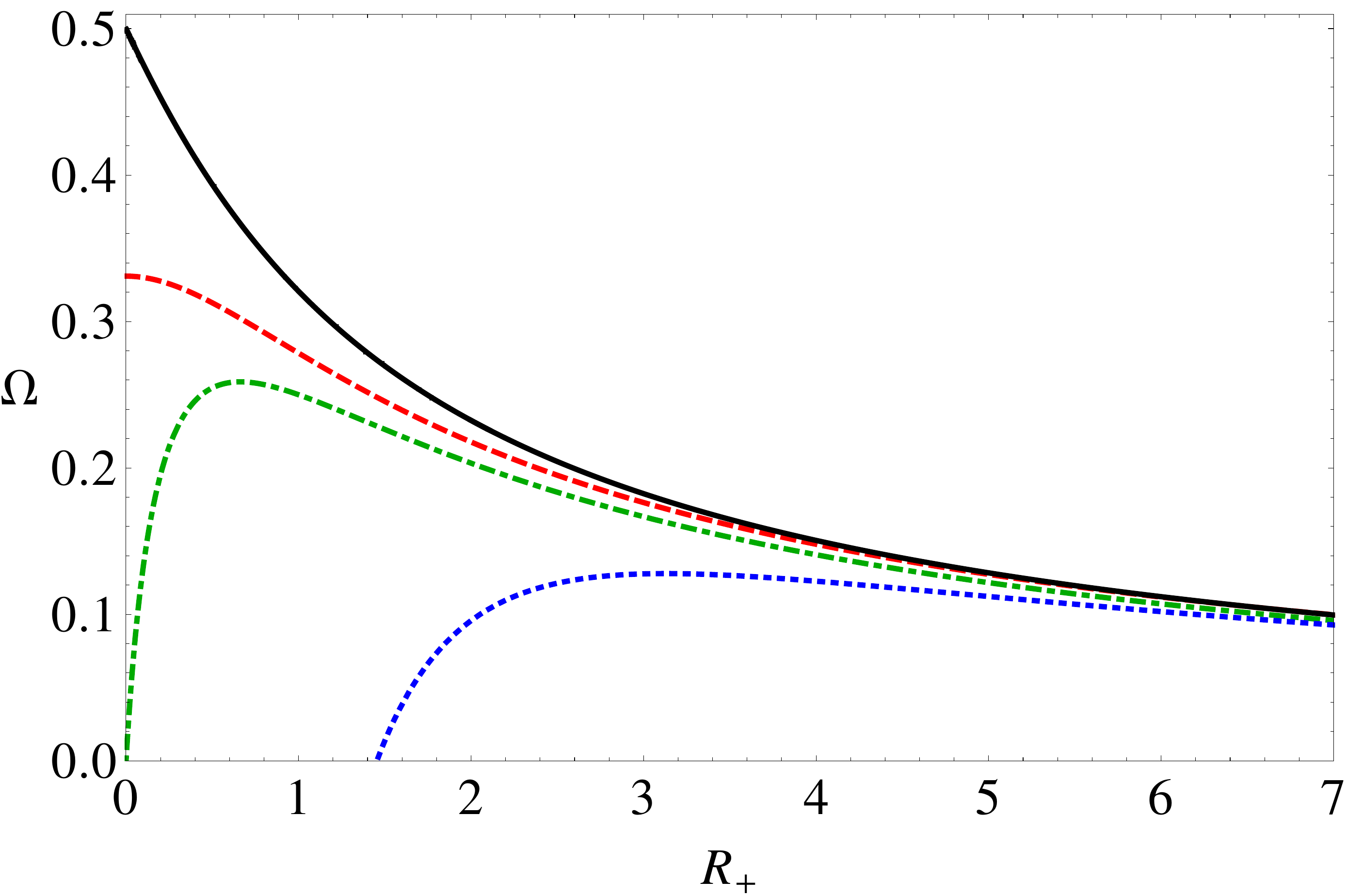} \caption{\label{fig:OmegaOverR+}Rotation frequency $\Omega$ versus hole radius
$R_{+}$ for spiral waves pinned to a Neumann hole. The analytical
result (black solid line, Eq. \eqref{eq:OmegaOverR+}) is compared
to Keener's result (green dot-dashed line \cite{tyson1988singular}),
the result by Hakim and Karma (blue dotted line \cite{hakim1999theory}),
and numerical solution of the linear eikonal equation (red dashed
line).}
\end{figure}
Because in general the difference between the core radius $r_{0}$
and the radius of the corresponding Neumann hole $r_{+}$ is small,
Keener's result can be applied to free spiral waves. This was done
successfully by Winfree, compare \cite{Winfree1991Alternative}. The
result obtained by Hakim and Karma \cite{hakim1999theory} can be
modified for pinned spirals according to 
\begin{align}
\omega_{\text{HK}}\left(r_{+}\right) & =\dfrac{c}{r_{+}}+\dfrac{2^{1/3}c^{1/3}a_{1}}{r_{+}^{5/3}}\epsilon^{2/3},\label{eq:OmegaHakimKarmaPinned}
\end{align}
where here $a_{1}=-1.01879$ denotes the global maximum of the Airy
function $\text{Ai}\left(x\right)$. The rescaled form of this expression
\begin{align}
\Omega_{\text{HK}}\left(R_{+}\right) & =\dfrac{1}{R_{+}}+2^{1/3}\dfrac{a_{1}}{R_{+}^{5/3}},
\end{align}
together with our result for $\Omega\left(R_{+}\right)$ and Keener's
result $\Omega_{\text{K}}\left(R_{+}\right)$, is compared in Fig.
\ref{fig:OmegaOverR+} with numerical simulations of the rescaled
eikonal equations Eqs. \eqref{eq:RescaledEikonalFront}, \eqref{eq:RescaledEikonalBack}.
All three analytical results agree well with the numerically obtained
curve in the limit of large hole radius $R_{+}\rightarrow\infty$
while for small core radii $R_{+}\rightarrow0$ partially marked deviations
appear. Merely our analytical approximation produces a finite and
different from zero rotation frequency for vanishing hole radius.
Note the different exponents in leading order of $\epsilon$ in the
expansions Eq. \eqref{eq:OmegaSmallEps}, Eq. \eqref{eq:OmegaSmallEpsKeener},
and Eq. \eqref{eq:OmegaHakimKarmaPinned}. 
\begin{figure}
\includegraphics[scale=0.3]{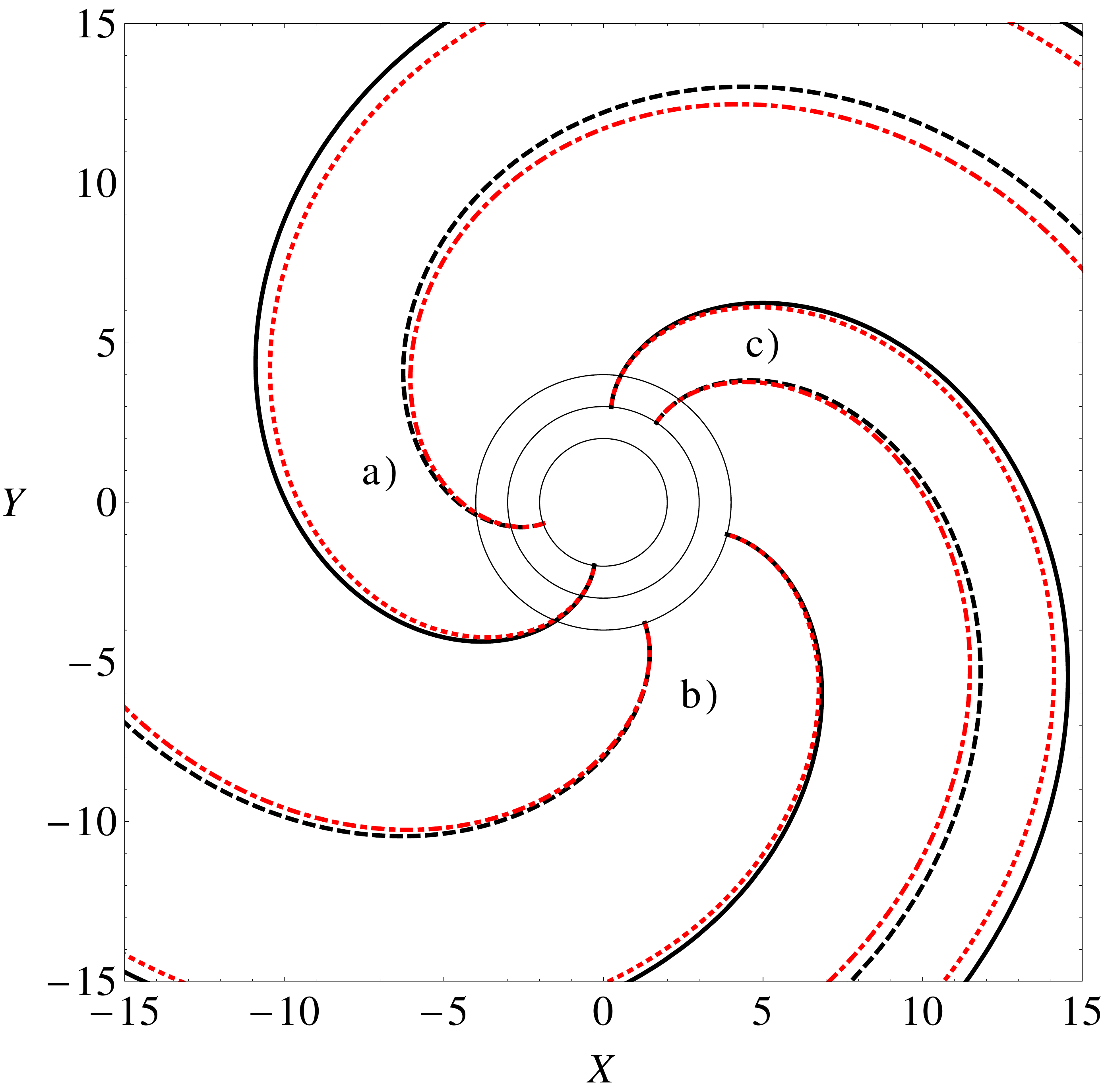}\caption{\label{fig:PinnedSpiralShape}Shape of spiral waves pinned to a Neumann
hole. Analytical result (black solid and dashed lines) and numerical
results (red dotted and dot-dashed lines) are plotted for different
hole radii and given excitability: a) $R_{+}=2,\, B=0.4,$ b) $R_{+}=4,\, B=0.3,$
c) $R_{+}=3,\, B=0.7.$}
\end{figure}

\subsection{Approximations for the wave shape $\Theta_{\text{ans}}^{\pm}\left(R\right)$}

\noindent From the Neumann boundary condition at $R=R_{+},$ we have
$R_{-}=R_{+}$ and it follows 
\begin{align}
\Psi_{\text{ans}}^{+}\left(R\right) & =\Psi_{\text{ans}}^{-}\left(R\right).
\end{align}
For the shape of the front interface we obtain 
\begin{align}
\Theta_{\text{ans}}^{+}\left(R\right) & =-\Omega\sqrt{R^{2}+R_{+}^{2}\left(\frac{4\Omega^{2}}{\left(R_{+}\Omega-1\right){}^{2}}-1\right)}\nonumber \\
 & +\frac{R_{+}\Omega\arccos\left(\frac{R_{+}}{R}\sqrt{1-\frac{4\Omega^{2}}{\left(R_{+}\Omega-1\right){}^{2}}}\right)}{\sqrt{1-\frac{4\Omega^{2}}{\left(R_{+}\Omega-1\right){}^{2}}}}.\label{eq:ShapePinnedSpiral}
\end{align}
The relation between hole radius $R_{+}$ and rotation frequency $\Omega$
is given by Eq. \eqref{eq:OmegaOverR+}. Furthermore, from the asymptotic
behavior of the solutions of the rescaled eikonal equations Eqs. \eqref{eq:RescaledEikonalFront},
\eqref{eq:RescaledEikonalBack} at $R=R_{+},$ we find $\Theta^{+}\left(R_{+}\right)-\Theta^{-}\left(R_{+}\right)=\dfrac{2\Omega}{B},$
from which follows 
\begin{align}
\Theta_{\text{ans}}^{+}\left(R\right) & =\Theta_{\text{ans}}^{-}\left(R\right)+\dfrac{2\Omega}{B}.
\end{align}
We compare the analytical predictions according to Eq. \eqref{eq:ShapePinnedSpiral}
with numerical solutions of the kinematic equations in Fig. \ref{fig:PinnedSpiralShape}.
Close to the hole, the agreement between theory and numerics is good.
With larger $R$ values, initially minor deviations arise which grow
according to 
\begin{align}
\Theta^{\pm}\left(R\right)-\Theta_{\text{ans}}^{\pm}\left(R\right) & =\gamma R+o\left(R\right),\; R\rightarrow\infty,
\end{align}
where $\gamma$ is some nonzero constant. The reason for this discrepancy
is that, first, the analytical relation $\Omega\left(R_{+}\right)$
is only approximately valid, and, second, the correct leading order
asymptotics for the shape functions is given by $\Theta^{\pm}\left(R\right)=-\Omega R+\mathcal{O}\left(\log\left(R\right)\right),\; R\rightarrow\infty$.
Our ansatz reproduces only the leading order asymptotics and does
not contain the logarithmic asymptotics.

\section{\label{sec:Conclusions}Conclusions}

Based on a new non-perturbative ansatz Eq. \eqref{eq:AnsatzPsi},
we have presented analytical approximations for spiral wave solutions
to the linear eikonal equation. The approximate analytical solution
$\theta_{\text{ans}}^{\pm}\left(r\right)$ for a rigidly rotating
spiral wave displays the correct leading order asymptotic expansion
of the unknown exact solution $\theta^{\pm}\left(r\right)$ close
to the core ($r\rightarrow r_{0}$), at the radius of the effective
Neumann hole ($r\rightarrow r_{+},\,\theta^{+}\vspace{0cm}'\left(r\right)$)
and far from the core ($r\rightarrow\infty$). In addition, approximate
and exact solution have the same asymptotic expansion in the inflection
point located on the wave back ($r\rightarrow r_{1},\,\kappa\left(r_{1}\right)=0$).
The asymptotically correct treatment of these four essential regions
of a spiral wave is crucial for our ansatz.\\
For the front interface our ansatz works quite well. In particular,
the derived dependence between the rotation frequency $\Omega$ and
the core radius $R_{0}$ for rigidly rotating spiral waves, Eq. \eqref{eq:OmegaOverR0},
agrees well with numerical solutions of the linear eikonal equation.
In fact, for large and intermediate core radii the agreement between
theory and numerics is very close, moreover, even for small core radius
our ansatz produces a more than acceptable match with the numerical
results. An equally good analytical approximation for $\Omega\left(R_{0}\right)$,
which is globally valid for all spiral core radii $R_{0}$, to our
knowledge, does not exist. Rotation frequency and core radius are
accessible in experiments with the Belousov-Zhabotinsky reaction,
for example. The relation for $\Omega\left(R_{0}\right)$, Eq. \eqref{eq:OmegaOverR0},
supplemented with the dispersion relation for one-dimensional pulse
trains, can be checked in experiments.\\
The relation for $\omega\left(r_{0}\right)$ can only be given implicitly.
Using perturbation theory, it is impossible to derive a globally valid
explicit approximation for $\omega\left(r_{0}\right)$ starting from
Eq. \eqref{eq:omegaoverr0}. The reason is that the small parameter
$\epsilon$ by introducing rescaled quantities drops out of Eq. \eqref{eq:OmegaOverR0}.
This scaling gives a dominant balance. There is no scaling giving
a dominant balance which leads from Eq. \eqref{eq:omegaoverr0} to
an equation which is simpler than Eq. \eqref{eq:OmegaOverR0}. In
other words, sooner or later we will inevitably be faced with Eq.
\eqref{eq:omegaoverr0} in order to obtain a globally valid solution
for $\omega\left(r_{0}\right)$.\\
We believe that one encounters the same situation for the full free
boundary approach based on the linear eikonal equations Eqs. \eqref{eq:LinearEikonalFront},
\eqref{eq:LinearEikonalBack}. They can be transformed to the rescaled
eikonal equations Eqs. \eqref{eq:RescaledEikonalFront}, \eqref{eq:RescaledEikonalBack}
that do not contain the small parameter $\epsilon$ any more. Within
any perturbative approach to Eqs. \eqref{eq:LinearEikonalFront},
\eqref{eq:LinearEikonalBack}, which is based solely on the assumption
of small $\epsilon$, the rescaled eikonal equations Eqs. \eqref{eq:RescaledEikonalFront},
\eqref{eq:RescaledEikonalBack} must be solved. Our relatively simple
analytical approximations might be a contribution to that approach.\\
Our ansatz gives simple analytical expressions for the shape of free
spirals, Eq. \eqref{eq:FreeSpiralShape}. Not only is the region far
from the core correctly represented, as it is also achieved by the
Archimedean and the involute spiral, but also the tip region of a
free spiral is modeled in accordance with numerical simulations of
the eikonal equation. For the shape of the back interface and the
dependence of the rotation frequency $\Omega\left(B\right)$ on the
excitability parameter $B$, the results produced by the ansatz are
much less satisfactory. The reasons for the discrepancy between analytical
prediction and numerical results are discussed in sections \ref{sub:OmegaOverB}
and \ref{sub:AnalyticalApproximationForTheSpiralShape}.\\
The relatively simple $R$-dependence for the wave shape Eq. \eqref{eq:FreeSpiralShape}
could be the starting point for stability analysis and further analytical
investigations.\\
In principle, one can improve the ansatz Eq. \eqref{eq:AnsatzTheta}
by including terms involving additional constants. Higher order asymptotics
at $r_{0},\, r_{+},\, r_{1}$ and for $r\rightarrow\infty$ can be
taken into account to determine these constants. However, it is difficult
to find terms which show the correct asymptotic behavior at one point
without simultaneously destroying the correct asymptotics at other
points.\\

\begin{acknowledgments}
We acknowledge support by the DFG via GRK 1558 (J. L.) and SFB 910
(H. E.).
\end{acknowledgments}
\bibliographystyle{apsrev4-1}
\bibliography{literature}

\end{document}